\documentclass[epj]{svjour}
\usepackage{graphicx}
\usepackage{amsmath}
\usepackage{amssymb}
\usepackage{amsfonts}
\begin{document}
\title{Developmental time windows for spatial growth generate multiple-cluster small-world networks}
\author{Florian Nisbach\inst{1} \and Marcus Kaiser\inst{2}% etc
}                     % Do not remove
\authorrunning{F. Nisbach and M. Kaiser}
\institute{University of Karlsruhe, Institut f\"{u}r Algebra und Geometrie, Englerstr. 2, 76128 Karlsruhe, Germany \and School of Computing Science, Newcastle University, Newcastle upon Tyne, NE1 7RU, U.K. \email{M.Kaiser@newcastle.ac.uk}}
\date{Received: date / Revised version: date}
\abstract{
Many networks extent in space, may it be metric (e.g. geographic) or non-metric (ordinal). Spatial network growth, which depends on the distance between nodes, can generate a wide range of topologies from small-world to linear scale-free networks. However, networks often lacked multiple clusters or communities. Multiple clusters can be generated, however, if there are time windows during development. Time windows ensure that regions of the network develop connections at different points in time. This novel approach could generate small-world but not scale-free networks. The resulting topology depended critically on the overlap of time windows as well as on the position of pioneer nodes.
 % end of PACS codes
} %end of abstract
\titlerunning{Developmental time-windows for spatial growth generate multiple clusters}
\authorrunning{F. Nisbach and M. Kaiser}
\maketitle
\section{Introduction}\label{sec:simulation}
Spatial networks often display high neighborhood connectivity---features of small-world networks \cite{Watts1998}---on the local and multiple clusters at the global level of organization. Clusters indicate sets of nodes with a higher proportion of edges within the same cluster than between different clusters. Note, that small-world networks may or may not consist of multiple clusters. For example, small-world networks generated by rewiring regular networks as in \cite{Watts1998} do not show multiple clusters. 

Whereas methods leading to small-world connectivity have been described before, there has been no systematic evaluation how multiple clusters arise during the growth of spatial networks. For example, in \cite{Kaiser2004b} we described spatial growth where the establishment of connections depended on the spatial distance between nodes. Whereas this approach yielded a wide range of topologies from small-world to linear scale-free, there was no guaranty that multiple clusters or communities would arise. 

Here, we show that multiple clusters can be reliably generated following the concept of time windows. Many real-world networks show an underlying time window structure -- that is, several parts of the network develop in certain stages of network growth, which are called their associated time windows. This is a well-known fact for the development of neural and cortical networks \cite{Sur2001,Rakic2002,Kaiser2007NC}, or for the evolution of non-spatial networks such as social networks, the Internet, or the World-Wide-Web.

We describe spatial growth with time windows in three dimensions in analogy to brain networks. However, similar considerations hold for two-dimensional spatial networks such as highway transportation networks \cite{Kaiser2004d}, the Internet \cite{Huberman1999,Yook2002}, or social networks \cite{Granovetter1973,Girvan2002}.

\section{Methods}
In our algorithm, the establishment of an edge depends on the distance between the nodes \cite{Kaiser2004b,Waxman1988,Kleinberg2000} and the current likelihood of establishing a connection given by the time windows of both nodes. The distance-dependent probability is 
\begin{equation}
P_{\text{dist}} = \beta \ e^{-\gamma \ d}
\end{equation}
where $d$ is the spatial Euclidean distance between two nodes, $\gamma=6$, and $\beta=6$. The effect of varying $\gamma$, and $\beta$ has been described previously \cite{Kaiser2004b}.

\begin{figure}[htbp]
	\centering
		\includegraphics[width=8.5cm]{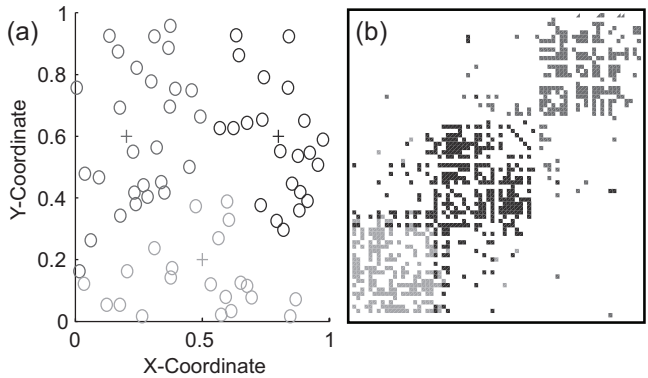}
	\caption{(a) Two-dimensional projection of three-dimensional node positions. The grey levels represent the time window corresponding to one of the three seed nodes (+).  (b) Adjacency matrix after the clustering algorithm has been applied showing three clusters. }
	\label{fig:seed}
\end{figure}

The time-dependent probability $P_{\text{time}}$ of a node is influenced by its distance to pioneer nodes ($N_1\ldots N_k\in\mathbb{R}^3$ where $k\in \mathbb{N}$ is the desired number of time windows). The reasoning is that nodes originate from other nodes in the region thereby inheriting their time domains from previous nodes (Fig.~\ref{fig:seed}a). These regions are the basis for network clusters (Fig.~\ref{fig:seed}b). Each node has a preferred time for connection establishment and the probability decays with the temporal distance to that time (see appendix \ref{sec:twf}, Fig.~\ref{fig:ptime1}).

A highly symmetric placement of the pioneer nodes is desirable. Therefore, their coordinates are chosen deterministically (i.e. only depending on $k$) in the following way: When it is possible, they are placed at the vertices of a platonic polyhedron (that is, for $k=4,6,8$), and otherwise equidistantly on a circle.

As the time $t$ in the simulation progresses from $0$ to $1$, new nodes and edges are added to the network in the following manner:
(1) Coordinates for a new node $U$ are chosen randomly (maximal distance to the origin is two units).
(2) $U$ is associated to the same time window as the nearest pioneer node, denoted by $w(U)$. 
(3) To each existing node $V$, a new edge is established with probability \begin{equation}\label{eqn:algorithmP}P=P_{\text{dist}}(d(U,V))\cdot P_{\text{time}}^{(w(V))}(t)\cdot P_{\text{time}}^{(w(U))}(t)
\end{equation}
where $d(U,V)$ denotes the Euclidean distance between the nodes $U$ and $V$.
(4)	If no edge to any existing nodes can be established, the new node is discarded. In that case, the time remains unchanged and steps (1)-(4) are repeated until a non-isolated node appears. This procedure ensures that the number of nodes grows linearly with time and that the final network contains the desired number of nodes.

With this implementation, the network topology was studied under the variation of the final number of nodes from 50 to 400 in steps of 50, the number of time windows from 3 to 10, and the "integral parameter" $\alpha$ of the time window functions, indicating the width of the time window (see appendix \ref{sec:twf}), from 0.1 to 0.6 in steps of 0.05. For each of the resulting parameter vectors, 40 networks were generated. The mean as well as the standard error of the mean was stored for each parameter vector. For simulations, we used MATLAB (Release 14, MathWorks Inc., Natick). Routines are available online (http://www.biological-networks.org).

\section{Results}
Significant correlations ($R^2>0.95$) of the measured values to the three parameters are listed and a functional correlation $\underline{\text{value}}\approx f(\underline{\text{parameter}})$ will be given. This relation means that the arithmetic mean over the parameter space of the other two parameters is calculated, and the resulting data is used for a correlation analysis.

\subsubsection*{Clustering coefficient (CC)}
\begin{figure}[htbp]
	\begin{center}
		\includegraphics[width=7cm]{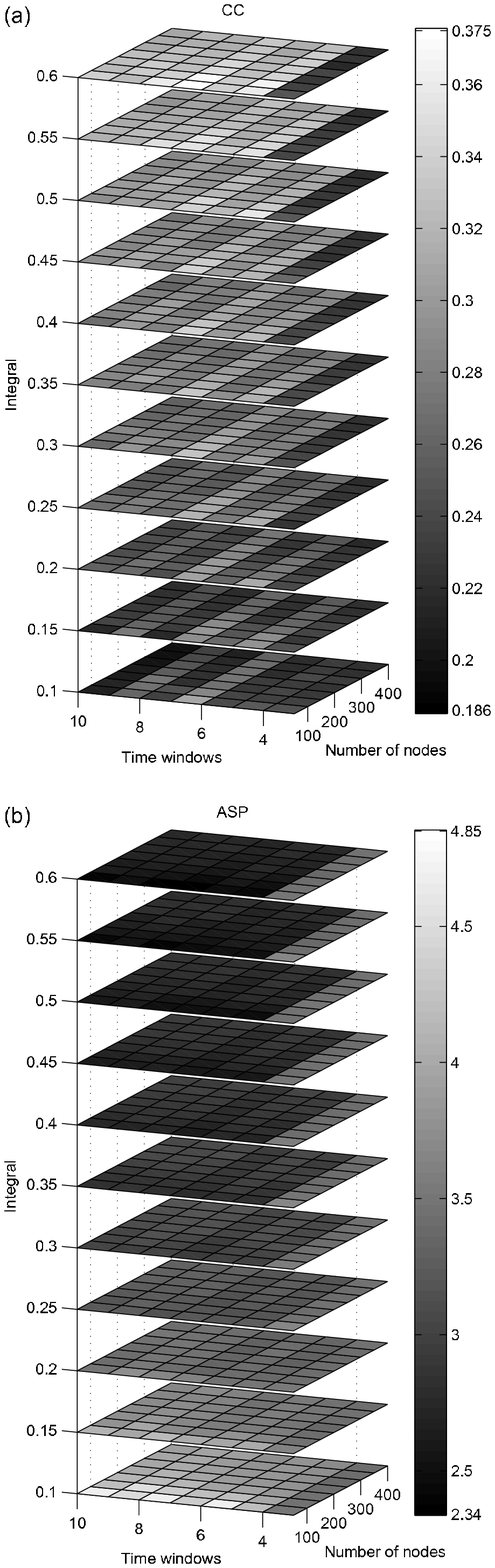}
	\end{center}

	\caption{(a) Clustering coefficient. (b) Average shortest path length.}
	\label{fig:CC}
	\label{fig:ASP}
\end{figure}

The clustering coefficient, calculated according to \cite{Watts1998}, ranges from 0.19 to 0.38 (Fig.~\ref{fig:CC}a). For three time windows, the CC is slightly negatively correlated with the number of nodes and almost independent from $\alpha$, varying between 0.21 and 0.26. For the other cases, CC positively correlates with $\alpha$ and negatively correlates with the number of nodes. 

\subsubsection*{Average shortest path (ASP)}
For three time windows, the ASP of the generated networks varies slightly around a mean of 3.45, almost independently of the other parameters. For more than three time windows, the ASP is nearly independent of the number of time windows, yet negatively correlated with the number of nodes, and $\alpha$ ($\text{ASP}\approx 2.33\,\alpha^{-0.25}\,$), varying in a range between 2.3 and 4.9 (Fig.~\ref{fig:CC}b).

\subsubsection*{Degree distribution}
\begin{figure}[htbp]
	\begin{center}
		\includegraphics[width=7cm]{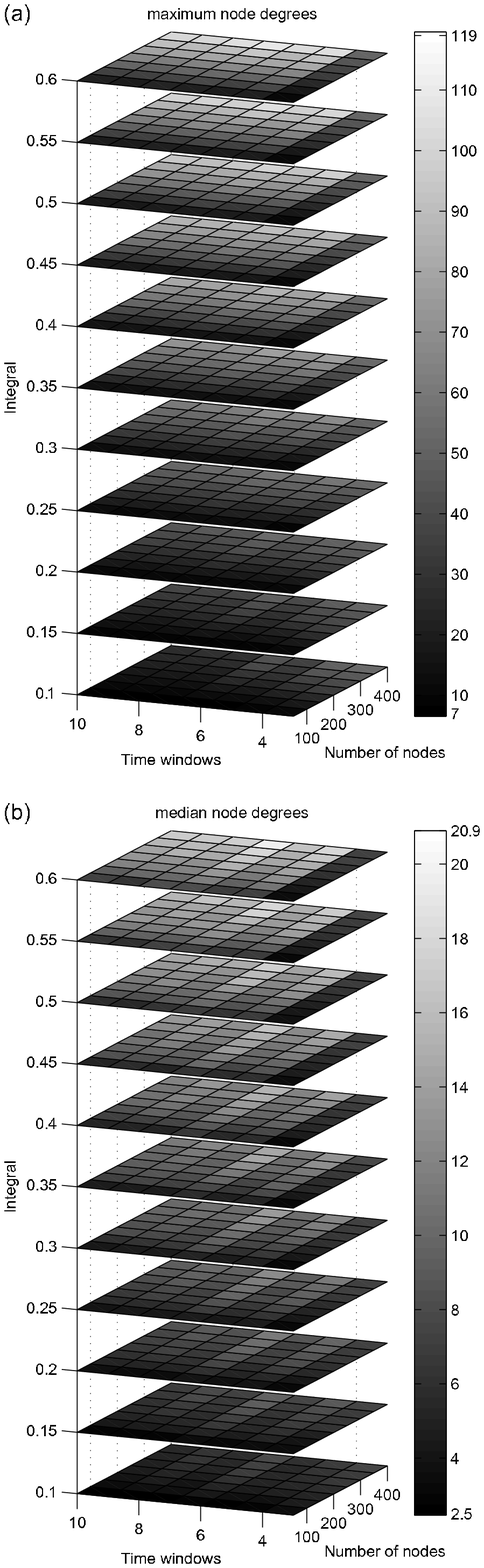}
	\end{center}

	\caption{Maximum and median degrees of nodes}
	\label{fig:dd}
\end{figure}

For all cases, the degree distribution does not follow a power law indicating that this model might not be able to produce scale-free networks \cite{Barabasi1999}. Figure \ref{fig:dd} shows plots of the maximum and median values of the node degrees. The maxima are linearly correlated with the number of nodes. They are also linearly correlated to the integral for more than three time windows.  The median of the degrees is correlated by a power law to the number of nodes with $y\approx 0.69\,x^{0.40}$ in the case of three time windows, and $y\approx 0.98\,x^{0.43}$ in the other cases.

\subsubsection*{Edge density and multiple clusters}
For three time windows, the number of edges is only positively correlated to the number of nodes (as expected, the number of edges grows like the square of the number of nodes), ranging from 90 to 2550. For the other cases, the number of edges ranges from 70 to 6550 and is positively correlated with the number of nodes $N$ (square), as well as with $\alpha$ (linearly). 

\begin{figure}[htbp]
	\begin{center}
		\includegraphics[width=7cm]{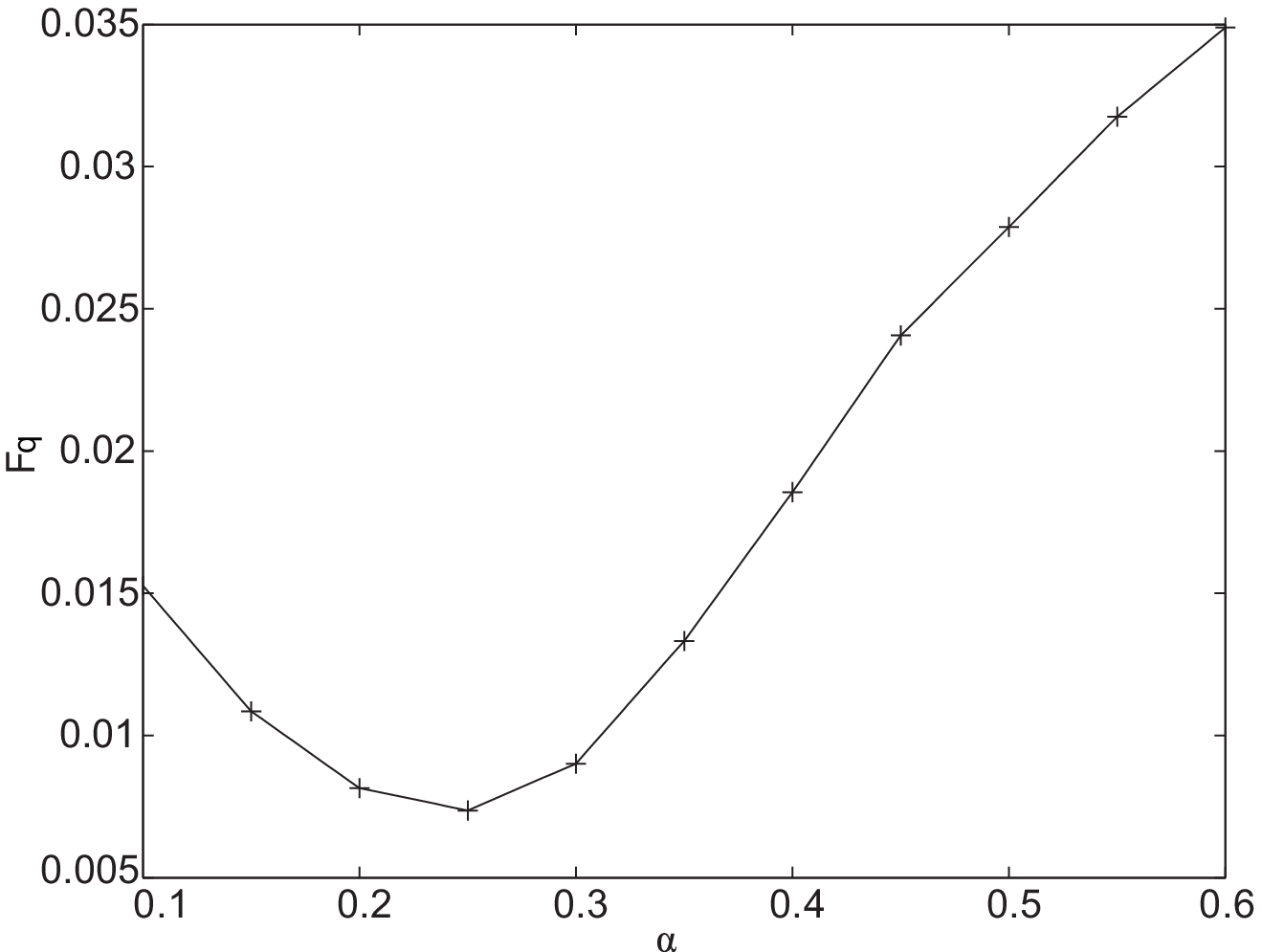}
	\end{center}

	\caption{Mean value of cluster connectedness $F_q$ for more than three time windows, plotted over integral size $\alpha$.}
	\label{fig:FQE_integral}
\end{figure}

As the generated networks show multiple clusters, it is important to know how densely the clusters are connected to each other. To get a single value still describing this interconnectedness, a measurement of "connections between the extremes" was introduced. Namely, for each simulation, a clustering algorithm (see appendix \ref{sec:clusteringalg}) was applied to the adjacency matrix. The interconnectivity was then measured as the ratio of edges that connect the first and last quartile of nodes in the clustered adjacency matrix ($F_q$). This indicates how well the first and the last cluster of a network were connected. For the case of three time windows, this cluster interconnectivity was very low, between 1.0\% and 1.9\%, showing a small negative correlation with the number of nodes. For the case of more than three time windows, there is a clear negative correlation with the number of nodes ($F_q \approx 1/N+C$). The constant $C$ depends on the range of $\alpha$ and the number of time windows over which the mean is calculated, but is significantly larger than zero (for the parameter ranges used here, we get on average $C \approx 0.16$). As shown in Fig. \ref{fig:FQE_integral}, the mean value of $F_q$ varies with $\alpha$, showing a local minimum for $\alpha=0.25$.

\subsubsection*{Small world properties}
\begin{figure}[htbp]
	\begin{center}
		\includegraphics[width=7cm]{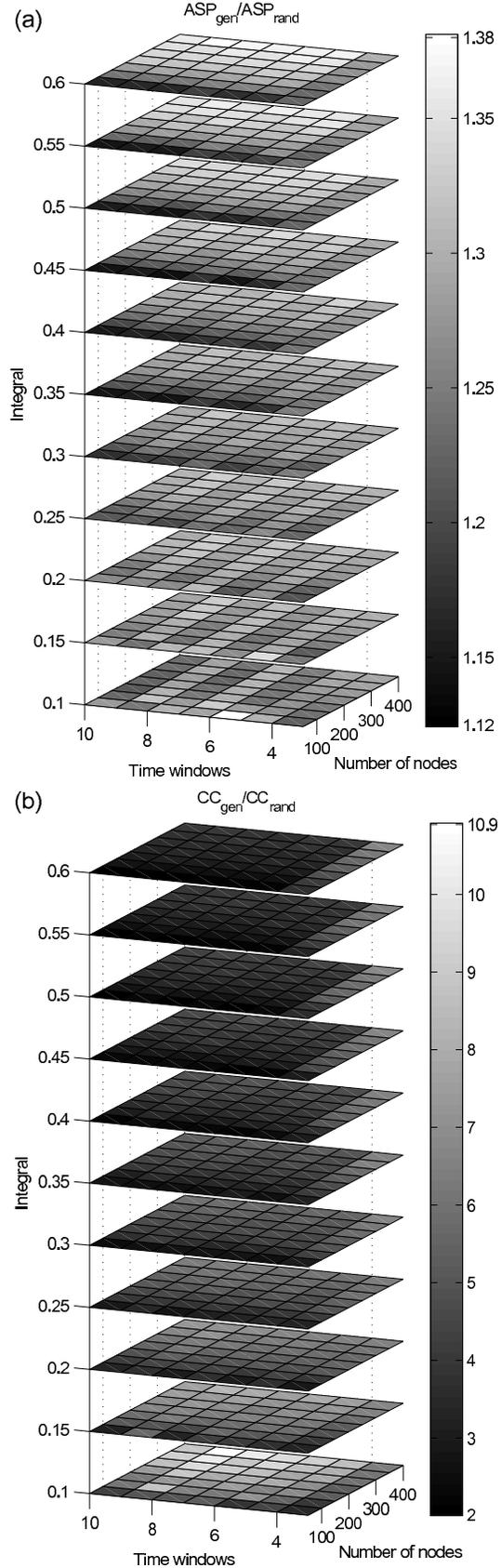}
	\end{center}

	\caption{$\text{ASP}$ and $\text{CC}$ ratios of the generated spatial networks relative to comparable random networks.}
	\label{fig:SW_ASP_CC}
\end{figure}

\begin{figure}[htbp]
	\begin{center}
		\includegraphics[width=7cm]{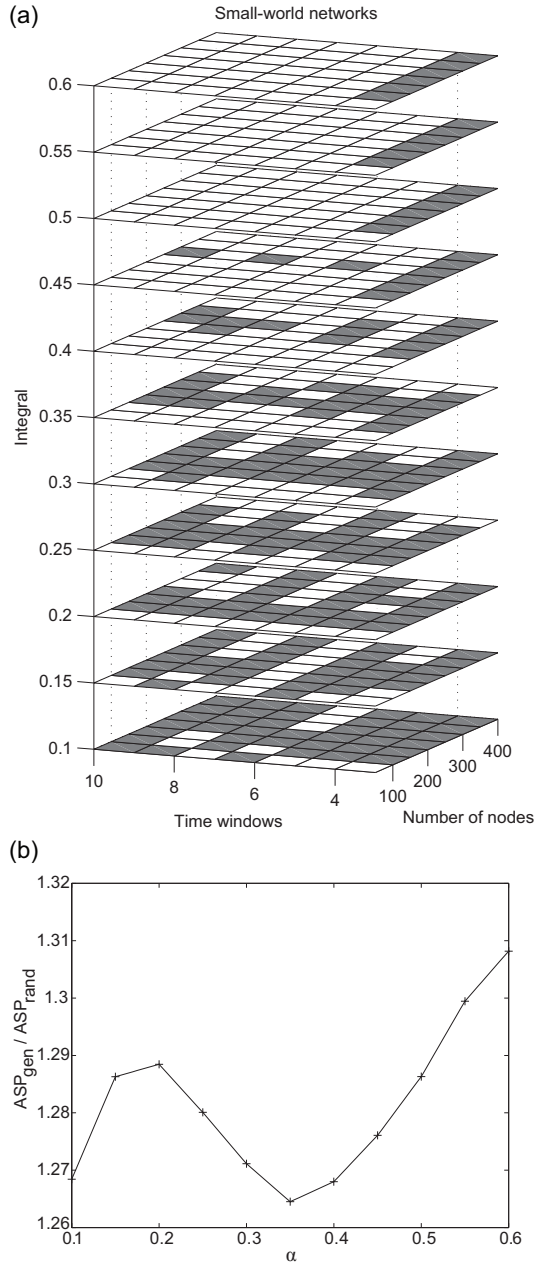}
	\end{center}

	\caption{(a) Networks of small world type (gray) (threshold: $\text{ASP}_\text{gen}/\text{ASP}_\text{rand}< 1.3$ and $\text{CC}_\text{gen}/\text{CC}_\text{rand}>4$); (b) Correlation between $\alpha$ and $\text{ASP}_\text{gen}/\text{ASP}_\text{rand}$}
	\label{fig:Smallworld}
	\label{fig:SWASP_integral}
\end{figure}

We tested for small-world features by comparing the clustering coefficients and average shortest paths with those of random benchmark networks (Fig.~\ref{fig:SW_ASP_CC}). For each network generated by spatial growth, 20 random networks were generated, and the mean parameter values were compared. A network was considered to be small-world if $\text{ASP}_\text{gen}/\text{ASP}_\text{rand}< 1.3$ and $\text{CC}_\text{gen}/\text{CC}_\text{rand}>4$. For three time windows, small world properties remain stable when changing the number of nodes and $\alpha$. For a higher number of time windows, the small-world feature is negatively correlated with $\alpha$ (Fig. \ref{fig:SWASP_integral}a). We have  $\text{CC}_\text{gen}/\text{CC}_\text{rand}\approx 0.75\,N^{0.35}$, $\text{CC}_\text{gen}/\text{CC}_\text{rand}\approx 2.6 / \sqrt{N}$,
$\text{ASP}_\text{gen}/\text{ASP}_\text{rand}\approx\sqrt{\alpha}$, and a correlation between $\alpha$ and $\text{ASP}_\text{gen}/\text{ASP}_\text{rand}$, with a local maximum for $\alpha=0.2$ and a local minimum at $\alpha=0.35$ (Fig. \ref{fig:SWASP_integral}b).

\subsubsection*{Time-based growth analysis} \label{ssec:timebased}
\begin{figure}[htbp]
	\begin{center}
		\includegraphics[width=7cm]{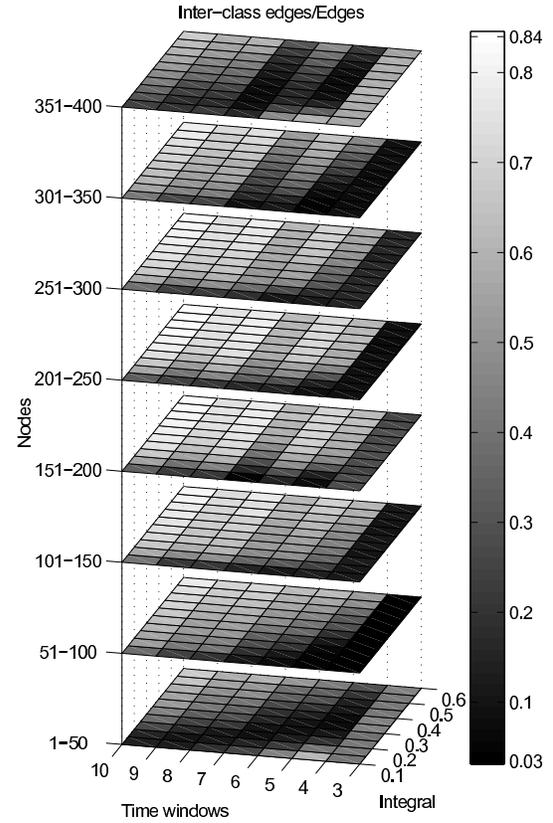}
	\end{center}

	\caption{Ratio of inter-class edges at different stages (number of generated nodes) of the network growth.}
	\label{fig:ratio_interedges}
\end{figure}

As our spatial networks grow over time, we also tested how network features evolved over time. For the same time window and $\alpha$ range as before, networks with 400 nodes were generated. For each step of 50 generated nodes, we observed the total number of edges as well as the number of inter-class edges between nodes with different time-window preferences for connection establishment. This tested whether there is a growth stage in which the generation of local (within clusters), or global (between clusters) connections occurs. In the results, again the three time window case is interesting: independently of $\alpha$, the majority of inter-class edges are established at the very beginning and the very end of growth, as well as in a very limited period when half of the final number of nodes were established (Fig. \ref{fig:ratio_interedges}). For more than three time windows---apart from some artifacts produced by pioneer node geometry---the inter-class growth did mainly occur around the half time of development.

\section{Discussion and conclusion}
The investigated approach was able to generate small-world multiple-cluster networks. We found that network topology is mainly influenced by the number of time windows and the spatial position of pioneer nodes and thus time domains.

The case of three time windows behaves remarkably different from the other numbers of time windows. A potential explanation could be that the low overlap of the time windows (see appendix \ref{sec:over}) allows only for a very slow network growth, in the sense that most nodes are discarded as they were not able to link to the existing network. 

Another critical parameter is the spatial position of pioneer nodes. For the connectivity within and between clusters, "artifacts" can be seen independently from the number of time windows. This gives evidence that these values depend on the actual placement of the pioneer nodes. Therefore, the interconnectivity and the size of network clusters could be adjusted by merely changing the placement of pioneer nodes while preserving the number of time windows.

In conclusion, we have presented a general framework for temporal growth of spatial networks depending both on the distance between nodes as well as on time domains for connection development. For neural systems, the time preference for establishing connections might be given by external/epigenetic factors in the surrounding medium or from genetic factors inherited from common precursor cells. Future work will have to investigate the role of pioneer node placement and mixing differently shaped, e.g. wide and narrow, time-windows.

\appendix
\section{Time window functions}\label{sec:twf}
The time window function had to meet the following criteria:
(1) The connection probability is zero at the start and end of development, i.e. $P_{\text{time}}^{(i)}(0)=P_{\text{time}}^{(i)}(1)=0$ for any $i$, 
(2) $P_{\text{time}}^{(i)}(\mu^{(i)})=1$, (3) $P_{\text{time}}^{(i)}$ should be continuously differentiable, and (4) In a set of time window functions, $P_{\text{time}}^{(i)}$, $i=1\ldots k$, the integral $I^{(i)}:=\int_0^1P_{\text{time}}^{(i)}(t)\:dt$ should be the same for any $i$. 	

\begin{figure}[htbp]
  \begin{center}
		\includegraphics[width=7cm]{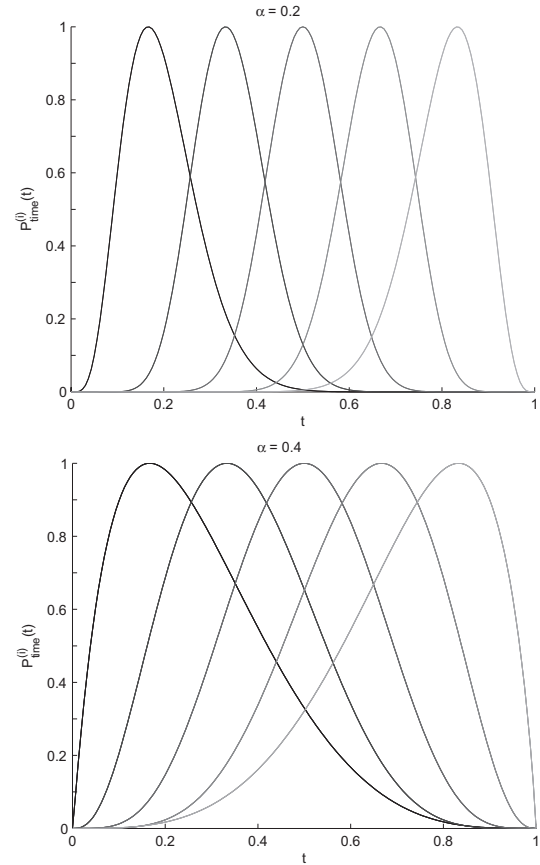}
	\end{center}
	\caption{Plot of $P_{\text{time}}^{(i)}$ for five time windows with (a) $\alpha=0.2$, or (b) $\alpha=0.4$.}
	\label{fig:ptime1}
	\label{fig:ptime2}
\end{figure}

These criteria were met for
\begin{equation}\label{eqn:twf}
P_{\text{time}}^{(i)}(t):=
P(t,\mu^{(i)},\Sigma_{\mu^{(i)}}(\alpha))=
\frac{1}{16}(t^{2\lambda}(t^\lambda-1)^2)^{\frac{1}{\Sigma_{\mu^{(i)}}(\alpha)}}
\end{equation}
with $\mu^{(i)}:=\frac{i}{k+1}$, $\lambda:=-\frac{\log(2)}{\log(\mu)}$ and $\alpha$ as the desired value of the integral, so that $\int_0^1 P_{\text{time}}^{(i)}(t)\: dt=\alpha $ (Fig. \ref{fig:ptime1}), and $\Sigma_\mu(\alpha)$ being a numerically determined scaling factor to get the desired integral value $\alpha$.

\section{The Overlap index} \label{sec:over}
For the time window functions given in equation \ref{eqn:twf}, or time window functions in general, the overlap between $P^{(i)}$ and $P^{(j)},\; i\neq j$, is, in some sense, a measure of how many connections are on average established between the correspondent node classes in a simulation. The definition of the overlap index $\Omega$ shall make this heuristic observation more precise. First of all, we will introduce the overlap $\text{Ov}(P^{(i)},P^{(j)})$ for two time window functions. Let us define
\begin{equation}
\begin{split}
\text{Ov}(P^{(i)},P^{(j)})&:=\frac{\int_0^1 P^{(i)}(t)P^{(j)}(t)\:dt}{\sqrt{\int_0^1 \left(P^{(i)}(t)\right)^2\:dt\cdot\int_0^1 \left(P^{(j)}(t)\right)^2\:dt}}\\
&=\frac{\langle P^{(i)},P^{(j)}\rangle }{\|P^{(i)}\|_2\cdot \|P^{(j)}\|_2}
\end{split}
\end{equation}
where $\langle\cdot,\cdot\rangle$ is the standard scalar product on the Euclidean vector space of integrable real valued functions on the interval $[0,1]$, and $\|\cdot\|_2$ is the associated norm. 

\begin{figure}[htbp]
	\begin{center}
		\includegraphics[width=8cm]{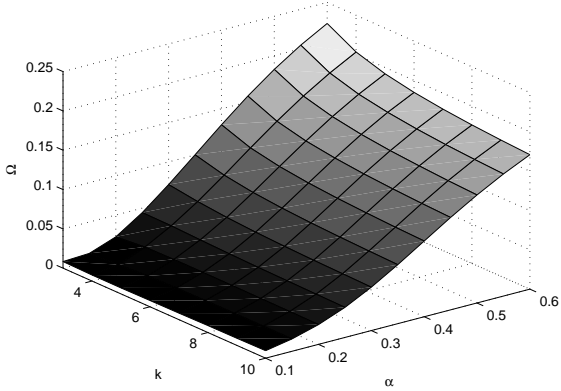}
	\end{center}
	\caption{Overlap index $\Omega$ of the standard time window functions depending on number of time windows, $k$, and the integral $\alpha$.}
	\label{fig:Omega}
\end{figure}

What we still need is a global measure $\Omega$ for the overlap, given a set $P^{(1)}\ldots P^{(k)}$ of time window functions used in a simulation. For this, we will take the average of  $\text{Ov}(P^{(i)},P^{(j)})$ for $i\neq j$, weighted by the
difference of time window indices $(j-i)$. This takes into account the fact that usually the maxima of the time window functions are placed at equal distance on the unit interval (remember we defined $\mu^{(i)}:=\frac{i}{k+1}$).
So we define
\begin{equation}
\Omega:=\frac{1}{S_k}\sum_{1\leq i < j \leq k} (j-i)\cdot\text{Ov}(P^{(i)},P^{(j)})
\end{equation}
What remains is to calculate the sum of the weights, $S_k$. This is done easily:
\begin{equation}
\begin{split}
S_k&=\sum_{1\leq i < j \leq k}(j-i)=\sum_{i=1}^k i(k-i)=k\sum_{i=1}^k i - \sum_{i=1}^k i^2\\
&=\frac{k^2(k-1)}{2}-\frac{k(k-1)(2k-1)}{6}
\end{split}
\end{equation}
So finally we get
\begin{equation}
\begin{split}
\Omega=&\frac{1}{\frac{k^2(k-1)}{2}-\frac{k(k-1)(2k-1)}{6}}\cdot\\
&\sum_{1\leq i < j \leq k} (j-i)\cdot\frac{\int_0^1 P^{(i)}(t)P^{(j)}(t)\:dt}{\sqrt{\int_0^1 \left(P^{(i)}(t)\right)^2\:dt\cdot\int_0^1 \left(P^{(j)}(t)\right)^2\:dt}}
\end{split}
\end{equation}
Figure \ref{fig:Omega} shows a plot of the Overlap index $\Omega$ for the case of the standard time window functions defined in equation \ref{eqn:twf}, where the number of time windows runs from $3$ to $10$, and the value of the integral, $\alpha$, runs from $0.1$ to $0.6$.

\section{The clustering algorithm} \label{sec:clusteringalg}
For the measurement of cluster interconnectivity $F_q$, a simple clustering algorithm was applied to the adjacency matrix. 

Given the adjacency matrix representation $(a_{ij})$ of a network, each element $a_{ij}$ stores a Boolean value saying whether the edge $(i,j)$ exists in the network:
\begin{equation}
a_{ij}= 
\begin{cases}
1 &\text{ nodes i and j are connected} \\
0 &\text{ otherwise}
\end{cases}
\end{equation}

We use a cosine-based similarity definition. 
Nodes are thought of as vectors in $n$ dimensional space, where $n$ is the number of nodes.  Node $i$ is represented by the $i$th column vector $\vec{i}$ of the adjacency matrix.
The similarity between two nodes is measured by computing the cosine of the angle between the associated vectors. 
The similarity between the nodes $i$ and $j$ is given by 
\begin{equation}m_{ij}=\cos \measuredangle(\vec{i},\vec{j})=\frac{\langle\vec{i},\vec{j}\rangle}{||\vec{i}||\cdot||\vec{j}||}\end{equation}
We call $m_{ij}$ the connectivity matching index between the nodes $i$ and $j$ ($i\neq j$).
It is a measurement for the similarity of their connection patterns. The entries $m_{ij}$ form
the connectivity matching matrix $M$.

Two nodes with a high matching index show a similar linking pattern. Therefore they are likely to be within the same cluster. 

The node $i$ with the highest number of connections to other nodes is chosen as initial node. It is the first node of the newly arranged network. The node  $j$ with the highest matching index $m_{ij}$ regarding node one is chosen to be node two. The node with the highest matching index regarding node two is labelled node three. The process continues until all nodes are labelled. A newly ordered adjacency matrix $A_C$ is then generated.

\begin{acknowledgement}
Marcus Kaiser was supported by EPSRC (EP/E002331/1). We thank Freya Gnam for developing the clustering algorithm described in appendix \ref{sec:clusteringalg}.
\end{acknowledgement}

%
% BibTeX users please use
%\bibliographystyle{harvard}
%\bibliography{ApsReport,neuro}
%

\newcommand{\noopsort}[1]{} \newcommand{\printfirst}[2]{#1}
  \newcommand{\singleletter}[1]{#1} \newcommand{\switchargs}[2]{#2#1}

\end{document}